\documentclass{elsart}

\usepackage{harvard}

 \usepackage{graphicx}
\usepackage{amssymb}
\def\la{\mathrel{\hbox{\rlap{\hbox{\lower4pt\hbox{$\sim$}}}\hbox{$<$}}}}
\def\ga{\mathrel{\hbox{\rlap{\hbox{\lower4pt\hbox{$\sim$}}}\hbox{$>$}}}}
\def\msun{M_\odot}
\def\astrobj#1{#1}
\newcommand{\bl}[1]{\mbox{\boldmath$ #1 $}}
\begin{document}

\begin{frontmatter}



\title{Radial transport of dust in spiral galaxies}


\author[UWO,IF]{E. I. Vorobyov},
\ead{vorobyov@astro.uwo.ca}
\author[RU]{Yu. A. Shchekinov}
\ead{yus@phys.rsu.ru}

\address[UWO]{Department of Physics and Astronomy, University of Western Ontario, London,
Ontario, N6A 3K7, Canada}
\address[IF]{Institute of Physics, Stachki 194, Rostov-on-Don, Russia }

\address[RU]{Department of Physics, Rostov State University, Rostov on Don,
           Russia}

\begin{abstract}
Motivated by recent observations which detect dust
at large galactocentric distances in the disks of spiral galaxies, we
propose a mechanism of outward radial transport of dust by spiral stellar
density waves. We consider spiral galaxies in which most of dust formation
is localized inside the corotation radius. We show that in the disks of such
spiral galaxies,
the dust grains can travel  over radial distances that exceed the corotation radius by roughly
$25\%$. A fraction of the dust grains can be trapped on kidney-shaped stable orbits
between the stellar spiral arms and thus can escape the destructive effect of
supernova explosions. These grains form diffuse dusty spiral arms, which
stretch 4-5~kpc from the sites of active star formation.
About $10\%$ of dust by mass injected
inside corotation, can be transported over radial distances $3-4$~kpc during $\approx 1.0$~Gyr.
This is roughly an order of magnitude more efficient than can be provided
by the turbulent motions.
\end{abstract}


\end{frontmatter}

% main text
\section{Introduction}
\label{}

Interstellar dust has been recently recognized to spread over large distances
from their source, and thus can be considered as a
tracer of dynamical processes responsible for the circulation of material in
disk galaxies. Observations of edge-on galaxies reveal a great amount of dust
far outside the galactic planes: in \astrobj{NGC 891} and \astrobj{NGC
4212} dusty clouds extend
up to $z\sim 2$ kpc above the plane, and contain in total as much as
$\sim 10^8~\msun$ (Howk and Savage, 1997, 1999; Rossa et al., 2004, Thompson
et al., 2004). Several mechanisms can contribute to transport of dust in the vertical
direction: convective motions or
bore flows \cite{Gomez} associated with spiral density waves,
chimneys produced by multiple supernova explosions \cite{Norman}, and
radiation pressure (Ferrara et al., 1991; Shustov and
Vibe, 1995; Dettmar et al., 2005).

The situation is less clear when the  radial distribution of dust is considered.
Observations of spiral galaxies in the far infrared band show that dust can extend in the radial direction
far beyond the stellar disks. Neininger et al. (\citeyear{Nein})
found that the radial distribution of 1.2 mm dust emission
from \astrobj{NGC 4565} is two times wider than CO emission. ISO observations of
several spiral galaxies at 200~$\mu$m (Trewhella et al., 2000; Bianchi
et al., 2000) revealed the presence of relatively
cold dust outside the stellar disks with a scale length $R_d\ga 1.5~R_\ast$
($R_d$ and $R_\ast$ are the radial scales of dust and stars, respectively).
A similar conclusion was made by Xilouris et al. (\citeyear{Xilouris}) for four spirals
observed in the optical and near infrared bands.
Bianchi et al. (\citeyear{Bianchi}) found that the best fit of the 200~$\mu$m emission
for \astrobj{NGC 6946} gives $R_d\ga 3~R_\ast$.
MHD turbulence can, in principle, spread dust outward in radial
direction \cite{Cho}, however, the corresponding time for galactic
scales seems to be too long. Indeed, the coefficient of turbulent diffusion
$D\sim \langle Lv\rangle/3$ is of the order of $10^{26}$ cm$^2$ s$^{-1}$ for typical interstellar
values $L\sim 100$ pc and $v\sim 10$ km s$^{-1}$. Therefore, dust
particles can diffuse by turbulent motions over scales $\sqrt{\langle \Delta
R^2\rangle}>3$ kpc in a characteristic time $t > 16$~Gyr.
In these conditions, only regular hydrodynamic motions can explain the presence of
dust outside stellar disks in spiral galaxies.

In this paper we explore the possibility that the radial transport of dust in
spiral galaxies is driven by the hydrodynamic motions associated with spiral
stellar density waves. In Sect.~\ref{model} a set of model hydrodynamic equations for dusty
interstellar medium is described and the equilibrium parameters are fixed.
In Sect.~\ref{spiral} the generation of spiral pattern in the stellar disk is formulated.
Sect.~\ref{test} describes trajectories of single dust particles
and Sect.~\ref{dust} describes the dynamics of dust fluid as a whole including
formation of dusty spiral arms. Sect.~6 summarizes our results.

\section{Model description}
\label{model}
\subsection{Basic equations}
\label{basic}
Our model disk is a two-fluid system composed of dust and gas, which
evolves in the external potential of the stellar disk and dark matter halo.
We assume that the dust and gas are coupled by friction. The gas disk is
isothermal at $T=10^4$~K and the dust pressure is negligible. We use the
thin disk approximation, i.e. all quantities are vertically integrated.
The system is described by the basic hydrodynamic equations,
where we denote the dust and
gas by the subscripts $d$ and $g$, respectively.

The continuity equation for the gas and dust is
\begin{equation}
{\partial \Sigma_{\rm g,d} \over \partial t}+ {\bl \nabla} \cdot (\Sigma_{\rm
g,d}\, {\bl v}_{\rm g,d})=0,
\label{cont}
\end{equation}
where $\Sigma_{\rm g,d}$ are the gas and dust surface densities and $\bl
v_{\rm d,g}$ are their velocities.

The equations of motion are
\begin{eqnarray}
{\partial {\bl v}_{\rm g} \over \partial t} +({\bl v}_{\rm g} \cdot {\bl
\nabla}) {\bl v}_{\rm g} &=& -{\bl \nabla} \Phi_{\rm s1,s2,h} - {{\bl \nabla} P_{\rm
g} \over \Sigma_{\rm g}} + {\Sigma_{\rm d} {\bl f} \over \Sigma_{\rm g}},
\label{first} \\
{\partial {\bl v}_{\rm d} \over \partial t} + ({\bl v}_{\rm d} \cdot {\bl
\nabla}) {\bl v}_{\rm d} &=& -{\bl \nabla} \Phi_{\rm s1,s2,h} -{\bl f},
\label{second}
\end{eqnarray}
where, $\Phi_{\rm s1}, \Phi_{\rm s2}$, and $\Phi_{\rm h}$ are the contributions
to the total gravitational potential by the axisymmetric stellar disk,
non-axisymmetric spiral density perturbation, and dark matter halo, respectively.
The friction force is defined as ${\bl f}=A ({\bl v}_{\rm d} - {\bl v}_{\rm
g})$ (see Draine and Salpeter, 1977; Noh et al., 1991).

We take the friction coefficient $A$ to be a simple function of the disk
conditions.
We assume that the thickness of the dust disk is comparable to that of the gas disk.
Then the collisional timescale $\tau_{\rm c}$ of energy and momentum exchange between
the gas and dust particles may be used as an estimate of $A^{-1}$,
\begin{equation}
A\sim \tau_{\rm c}^{-1}={\sigma_{\rm c} n_{\rm g} v_{\rm th} m_{\rm g} \over m_{\rm
d}},
\label{thick}
\end{equation}
where $n_{\rm g}$ is the number density of gas particles, $m_{\rm g}$ and
$m_{\rm d}$ are the masses of gas and dust particles, and $v_{\rm th}=\sqrt{
3RT/\mu}$ is the gas velocity dispersion; $\tau_{\rm c}$ is estimated for
the subsonic relative motion of neutral dust grains and gas.
Since we use the thin disk approximation, the gas volume density in Eq.~(\ref{thick})
is approximated assuming a local vertical hydrostatic equilibrium
$m_{\rm g} n_{\rm g}=\Sigma_{\rm g}\,/ z_0$,
where the gas disk thickness is defined as\footnote{Since
the dominant component of gravity in the vertical direction is the stellar disk,
the total surface density ($\Sigma_{\rm tot}=\Sigma_{\rm g} + \Sigma_{\rm s}$) should appear in the expression for $z_0$.
This may increase the friction coefficient $A$, especially in the inner
regions.  We discuss this issue in Sect.~\ref{dust}}
$z_0=2c_s^2/(\pi G \Sigma_{\rm g})$, where $c_{\rm s}$ is the sound speed.
The geometrical cross section is used as
an estimate of the collisional cross section $\sigma_{\rm c}=\pi a^2$,
where $a$ is the average radius of a dust particle.
In the present simulations, we assume $m_{\rm d}=10^{-14}$~g and
$a=10^{-5}$~cm. It is well known that interstellar dust grains are charged,
and Coulomb interactions with charged ambient ions and electrons increase
the friction force (Draine and Salpeter, 1979; Weingartner and Draine, 2001).
However, as we are mostly concerned with the diffuse HI phase of the ISM (warm neutral
medium, WNM, in standard nomenclature), this increase does not crucially change
 the resulting friction coefficient.
Indeed, the Coulomb friction can be written as
$F_{\rm C}=\chi(Z)F^0_{\rm HI}$, where
$F^0_{\rm HI}$ is the friction in an HI environment, $\chi(Z)=\langle
a_Z/a\rangle^2\ln\Lambda\sum\limits_i^{}x_i(m_i/m_p)$, $a_Z=Ze^2/kT$,
$\ln\Lambda$ is the Coulomb logarithm, $x_i$ is the fractional ionization of
$i$-th ions or electrons, $m_i$ is the corresponding mass, averaging is
over charge distribution of dust grains \cite{Weingartner}.
For typical conditions in the WNM; $x\sim 0.1$, $n\sim 0.3$
cm$^{-3}$, and dust charge $\langle Z\rangle\simeq 60$ \cite{Yan}
we get $\chi\sim 1$. Therefore, we will restrict our consideration with
the friction coefficient $A$ as determined for the neutral dust.

\subsection{Equilibrium gas disk}
\label{equilib}
Initially, our model galaxy consists of a rotating gas disk balanced
by the gravity of an {\it axially symmetric} stellar disk and spherically
symmetric dark matter (DM) halo.
The dust disk is introduced later when the non-axisymmetric part of the
stellar gravitational potential is set.
The rigid stellar disk is assumed to have a power-law radial density profile
of the form
\begin{equation}
\Sigma_{\rm s}(r)= {B^2 \over 2 \pi G} \left[(r_{\rm s}^2+r^2)^{-3/2} \right],
\label{stellar}
\end{equation}
where $B^2= 2 \pi G r_{\rm s}^3 \Sigma_{\rm s0}$.
The gravity force of such a density distribution
is given by Toomre (\citeyear{Toomre})
\begin{equation}
{\partial \Phi_{\rm s1} \over \partial r}=B^2\left[{r\over r_{\rm s}} (r_{\rm s}^2+r^2)^{-3/2}\right].
\label{sym}
\end{equation}
In the following, we use the central stellar density
$\Sigma_{\rm s0}=1.2\times 10^3~M_\odot$~pc$^{-2}$ and $r_{\rm s}=3$~kpc,
which gives us a total stellar mass of $M_{\rm st}=7\times 10^{10}~M_\odot$.

The density distribution of the rigid DM halo
is  assumed to be that of a modified isothermal sphere \cite{BT}
\begin{equation}
\rho_{\rm h}={\rho_{\rm h0}\over (1+r/r_{\rm h})^2},
\label{halodens}
\end{equation}
where the central density  $\rho_{\rm h0}$
and  the characteristic scale length $r_{\rm h}$ are given by Mac Low
and Ferrara (\citeyear{MF}) and Silich and Tenorio-Tagle (\citeyear{Silich}):
\begin{equation}
\rho_{\rm h0}=6.3 \times 10^{10} \left( {M_{\rm h} \over M_{\odot}}
\right)^{-1/3}~h^{-1/3}~M_{\odot}~{\rm kpc}^{-3}
\label{halocentr}
\end{equation}
\begin{equation}
 r_{\rm h}=0.89 \times 10^{-5} \left( {M_{\rm h} \over
 M_{\odot}}\right)^{1/2}~h^{1/2}~{\rm kpc}.
\label{halorad}
\end{equation}
Here, $h$ is the Hubble constant in units of 100~km~s$^{-1}$~Mpc$^{-1}$ and
$M_{\rm h}=10^{12}~M_\odot$ is the total DM halo mass. We note that the
DM halo mass within the computational domain $r<20$~kpc is $M_{\rm h}(r<20~{\rm
kpc})=5 \times 10^{10}~M_\odot$.
We adopt $h=0.65$ throughout the paper.
The gravitational force of spherically symmetric DM halo
can be expressed as
\begin{equation}
{\partial \Phi_{\rm h} \over \partial r}=4 \pi G \rho_{\rm h0}
r_{\rm h}\left[ r/r_{\rm h} - \arctan(r/r_{\rm h}) \right]
\left({r_{\rm h}\over r}\right)^2.
\label{halo}
\end{equation}

The gas disk has an exponentially declining density profile with
the central surface density $\Sigma_{\rm g0}=30~M_\odot$~pc$^{-2}$,
and radial scale length $r_{\rm g}=9$~kpc. The total mass of the gas disk
within the computational domain ($r=20$~kpc) is
$M_{\rm g}=1.0 \times 10^{10}~M_\odot$. The gas disk contains only
a small fraction ($\sim 10\%$) of the total mass in the computational domain.
Hence to a first approximation we can assume that the gas moves in the
potential field of the stars and DM halo.

\begin{figure}
  \resizebox{\hsize}{!}{\includegraphics{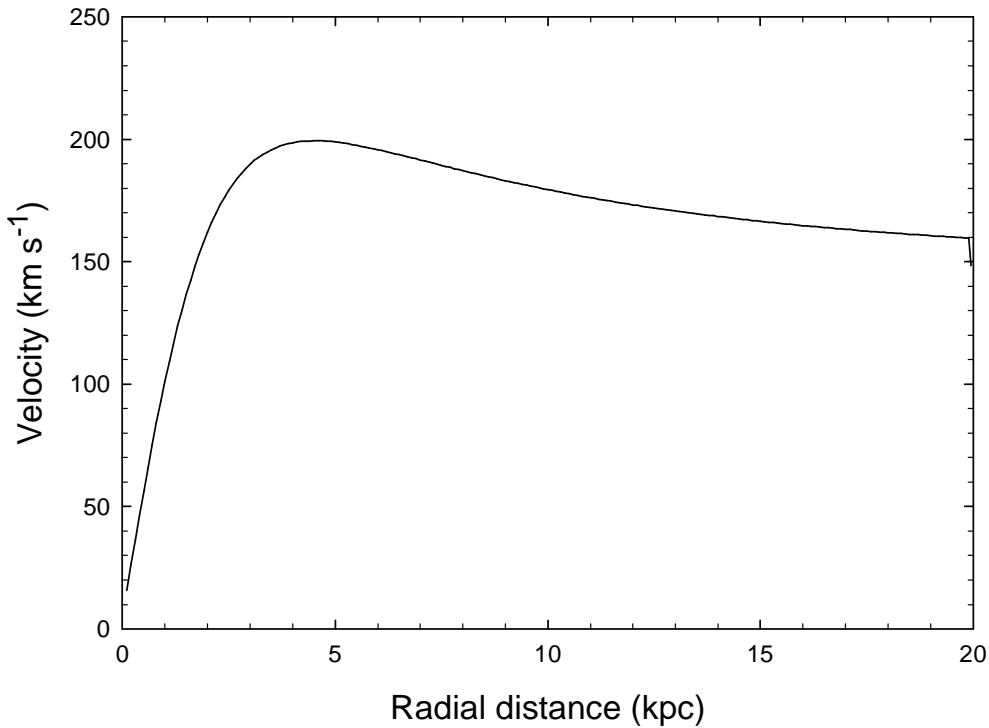}}
      \caption{The initial rotation curve of the gas disk.}
         \label{fig1}
\end{figure}

Once the density profile of the gas disk is fixed,
its initial rotation curve (RC) is obtained by solving Eq.~(\ref{first}), where the friction
force $f$ is set to zero and the time variations are neglected. The resulting initial RC is
plotted in Fig.~\ref{fig1}, where it reaches
a maximum circular velocity of 200~km~s$^{-1}$ at $r\approx 3$~kpc. Beyond
this maximum, the rotational velocity undergoes a gradual decline.
Because the gas surface density has a slowly declining radial profile ($r_{\rm
g}=9$~kpc) and the gas is assumed to be isothermal,
the contribution of the pressure gradient in the gas rotation curve is
less than $1\%$ everywhere in the disk.

\subsection{Computational techniques}
\label{comp}
A set of hydrodynamic equations (\ref{cont})-(\ref{second}) in polar coordinates ($r,\phi$)
is solved using the method of finite-differencing with a time-explicit,
operator-split solution procedure (ZEUS-2D) described in detail in Stone
and Norman (\citeyear{SN}). We use a resolution of $200\times200$ grid points in a
polar computational domain with a radius of 20~kpc.
We set a free boundary condition at the
outer boundary, i.e. the gas and dust are allowed to flow out freely
from the computational domain. The outflowing gas/dust is assumed to be lost
by the parent galaxy.

\section{Generation of spiral pattern}
\label{spiral}
\begin{figure}
  \resizebox{\hsize}{!}{\includegraphics{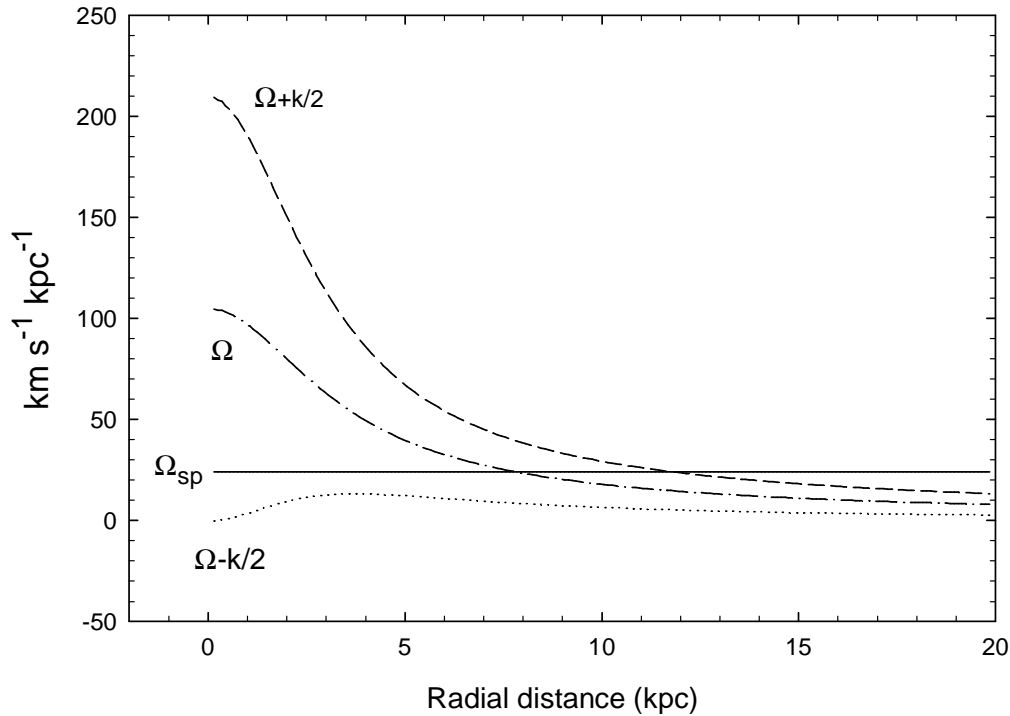}}
      \caption{Behaviour of $\Omega \pm \kappa/2$, where $\kappa$ is the
      epicyclic frequency and $\Omega$ is the angular velocity of gas.
      The angular velocity of spiral pattern $\Omega_{\rm
      sp}$ is shown by the solid line. }
         \label{fig2}
\end{figure}

Once the axisymmetric equilibrium gas disk is constructed, we introduce
the non-axisymmetric spiral gravitational potential of the stellar disk
$\Phi_{\rm s2}$.
The latter is defined in the form of a running spiral density wave \cite{Lin}
\begin{equation}
\Phi_{\rm s2}=-C(r) \cos\left[ m(\cot(i) \ln(r/r_{\rm sp})+\phi - \Omega_{\rm sp} t)\right].
\label{nonsym}
\end{equation}
Here, $C(r)$ is the amplitude of the spiral gravitational potential,
$m$ is the number of spiral arms,
$i$ is the pitch angle, $r_{\rm sp}$ is the characteristic winding length of
spiral arms, and $\Omega_{\rm sp}$ is the
angular velocity of spiral pattern.
In the following we adopt $m=2$, $i=25^{\circ}$,
$r_{\rm sp}=6$~kpc, $\Omega_{\rm sp}=24$~km~s$^{-1}$~kpc$^{-1}$. This choice
of $\Omega_{\rm sp}$
places the corotation of the gas disk at $\approx 8.0$~kpc and the outer
Lindblad resonance at $\approx 12.0$~kpc. The inner Lindblad resonance
is absent. The behavior of gas angular
velocity $\Omega$ and $\Omega \pm \kappa  /2$ is shown in Fig.~\ref{fig2}.
According to Kennicutt (\citeyear{Ken1}), a pitch angle of $i=25^\circ$
is typical for Sc galaxies. The efficiency of radial transport of
dust for different pitch angles is discussed in Sect.~\ref{dust}.

\begin{figure}
  \resizebox{\hsize}{!}{\includegraphics{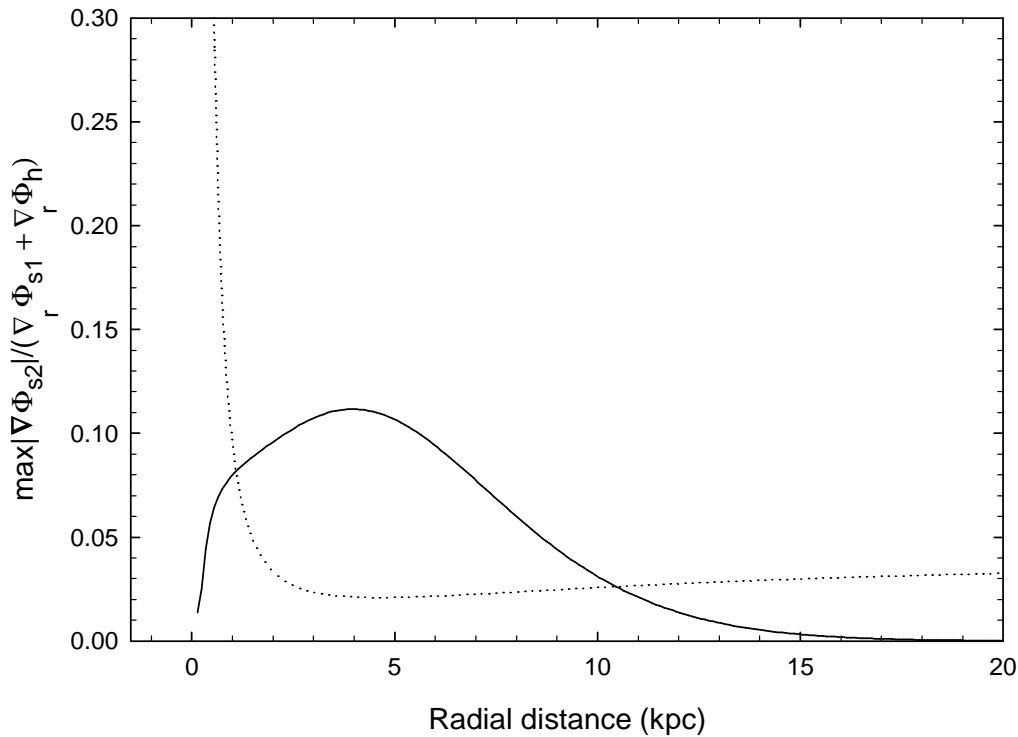}}
      \caption{ The ratio of the maximum, non-axisymmetric perturbing force $|{\bl\nabla} \Phi_{\rm
      s2}|=[(\partial \Phi_{\rm s2}/\partial r)^2+(r^{-1}\partial \Phi_{\rm s2}/\partial
      \phi)^2]^{1/2}$ to the total axisymmetric gravity force
      $(\partial \Phi_{\rm s1}/\partial r+ \partial\Phi_{\rm h}/\partial
      r)$  as a function of radial distance. The solid
      and dashed lines are explained in greater detail in the text.}
         \label{fig3}
\end{figure}

The logarithmic spiral defined by Eq.~(\ref{nonsym}) is quickly winding
up as $r\rightarrow0$. This may create large gradients
of gravitational potential near the origin and lead to serious numerical
difficulties. For instance, if the amplitude $C(r)$ is independent of $r$, than
the ratio of maximum non-axisymmetric perturbing force
to the total axisymmetric gravity force $\beta(r)=max|{\bl\nabla} \Phi_{\rm
s2}|/(\partial \Phi_{\rm s1}/\partial r+ \partial\Phi_{\rm h}/\partial
r)$ sharply increases as $r\rightarrow 0$.  This is illustrated
in Fig.~\ref{fig3}, where $\beta(r)$ is plotted by the dotted line for $C=0.005$
(in dimensionless units).
The apparent growth of $\beta$ with radius for $r>3$~kpc
is also unrealistic
since the spiral arms are expected to vanish  at larger radii. In view
of these difficulties, we use a radially dependent amplitude of the form
$C(r)=C_0(r)^{\alpha(r)}$. Here, $C_0(r)$ is a linear function of $r$ which has
a value of 0 at $r=0$~kpc (ensuring that  $|{\bl\nabla} \Phi_{\rm s2}|$ diminishes as $r\rightarrow 0$)
and attains its maximum value of 0.0045 (in dimensionless units) at $r=20$~kpc.
The exponent $\alpha(r)$ decreases linearly
with radius from $\alpha=2$ at r=0~kpc to $\alpha=-2.1$ at r=20~kpc.
The resulting profile of $\beta(r)$ is shown in Fig.~\ref{fig3}
by the solid line.
Since the non-axisymmetric gravity force scales
as $|{\bl\nabla} \Phi_{\rm s2}| \propto 1/r^{1-\alpha(r)}$, the ratio $\beta$
increases with radius at $r<4$~kpc, and decreases at $r\ga 4$~kpc.
At the position of corotation $r_{\rm cr}\approx 8$~kpc, $\beta$ drops by roughly a factor
of 2 as compared to its maximum value at $r\approx 4$~kpc and continues to decrease at larger radii.
We note that the maximum, non-axisymmetric
perturbing force never exceeds $12\%$ of the total axisymmetric gravity
force.

\begin{figure}
  \resizebox{\hsize}{!}{\includegraphics{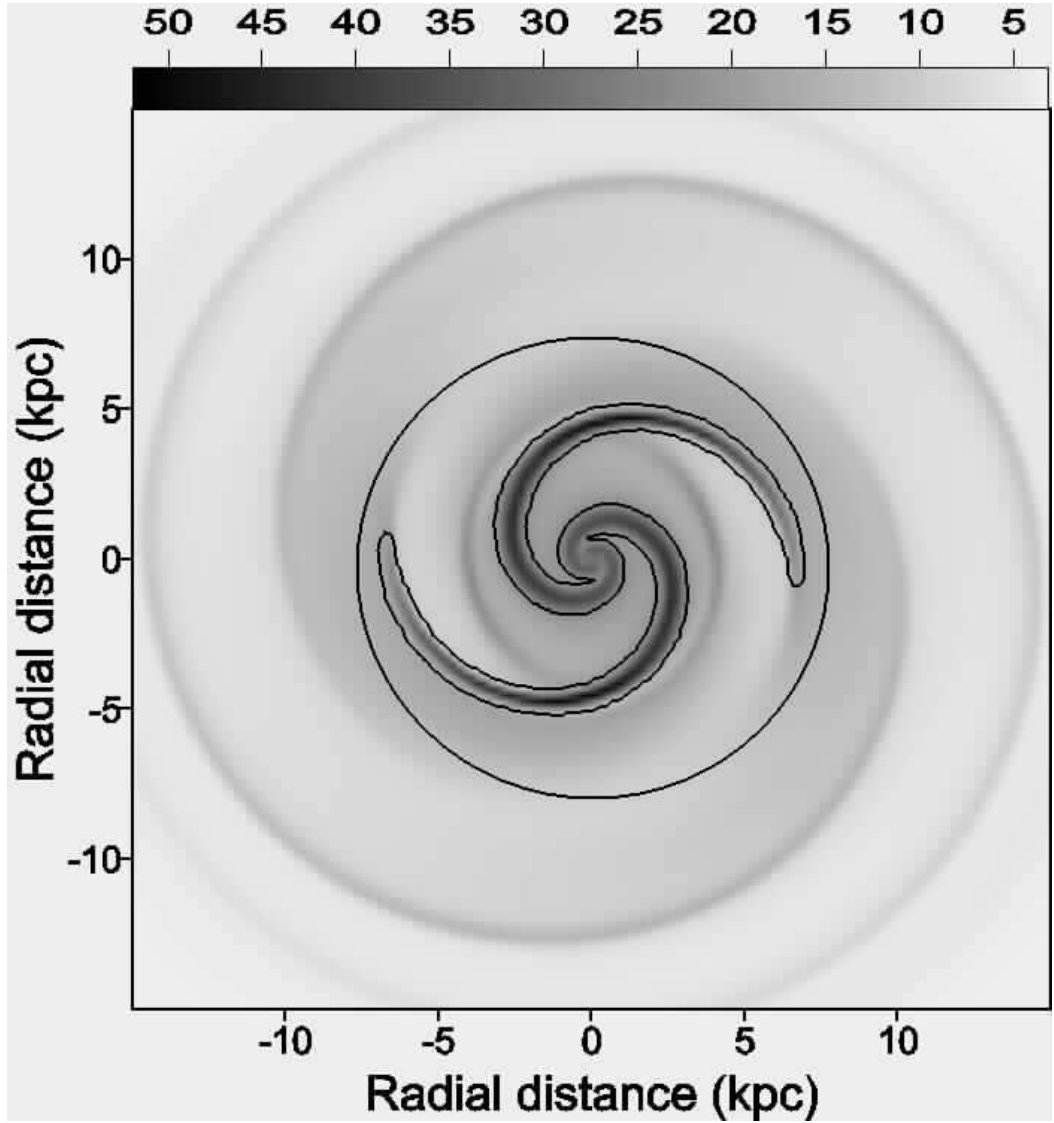}}
      \caption{Gas density distribution at $t=0.8$~Gyr from the beginning
      of numerical simulations. The corotation radius is shown by the
      circle. The scale bar is in $M_\odot$~pc$^{-2}$. The contour line
      sketches the region of supercritical gas density (see Sect.~\ref{dust} for
      details).}
         \label{fig4}
\end{figure}

The non-axisymmetric part of the stellar gravitational potential $\Phi_{\rm
s2}$ is turned on slowly. Specifically, $\Phi_{\rm s2}$ is multiplied by
a function $\epsilon(t)$, which has a value of $0$ at t=0 and linearly
grows to its maximum value of $1.0$ at $t\ge200$~Myr.
It takes a few hundred Myr for the gas disk to adjust to the spiral distortion
and develop a spiral structure as shown in Fig.~\ref{fig4}. The gas spiral pattern
(as well as the stellar one) rotates counterclockwise.
The position of corotation is shown by the circle.
The strongest gas response to the underlying stellar spiral density wave
is seen inside the corotation circle. However, a weak spiral structure
can still be traced outside corotation at radii $r=10-15$~kpc.

\begin{figure}
  \resizebox{\hsize}{!}{\includegraphics{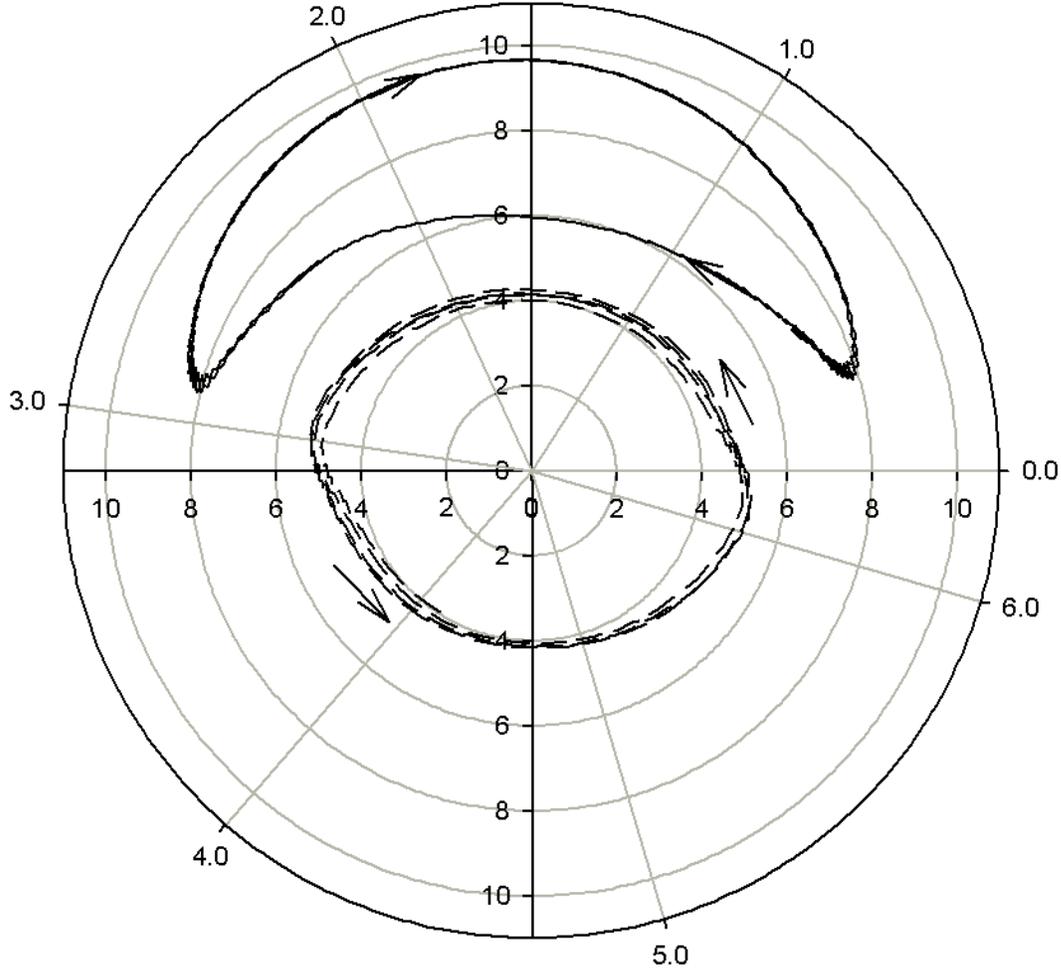}}
      \caption{Polar plot ($r, \theta$) showing trajectories of three test dust particles.
      The radial distance is in kiloparsecs and the azimuthal angle is in radians.
      Stellar spiral arms are located along azimuthal angles of 0 and $\pi$.
      The direction of motion of test particles in the frame of reference
      of stellar spiral arms is indicated by arrows.}
         \label{fig5}
\end{figure}

\section{Trajectories of dust particles}
\label{test}
In this section we study the trajectories of single dust particles in the
combined gravitational potential of stellar disk and dark matter halo
including the effects of dust-gas friction.
The equation of motion of the
dust fluid~(\ref{second}) can be obviously reduced to a set of equations
of motion for a single dust particle and subsequently
solved using the Runge-Kutta scheme. We introduce
test dust particles at different radial distances from the galactic centre
$r_{\rm d}$ and at different azimuthal angles $\theta_{\rm d}$
with respect to the position of stellar spiral arms. In our simulations, the stellar spiral arms
are modeled by the spiral gravitational potential rotating counterclockwise
in the laboratory frame of reference.
Therefore, the position of stellar spiral arms at any given time
is determined as the position of minimum in the spiral
gravitational potential defined by Eq.~(\ref{nonsym}).
All dust particles are injected at $t=0.8$~Gyr when the gas spiral structure is fully
developed (see Fig.~\ref{fig4}), and they are assigned velocities drawn from
the parent gas. We followed the trajectories of test dust particles for 2~Gyr.

\begin{figure}
  \resizebox{\hsize}{!}{\includegraphics{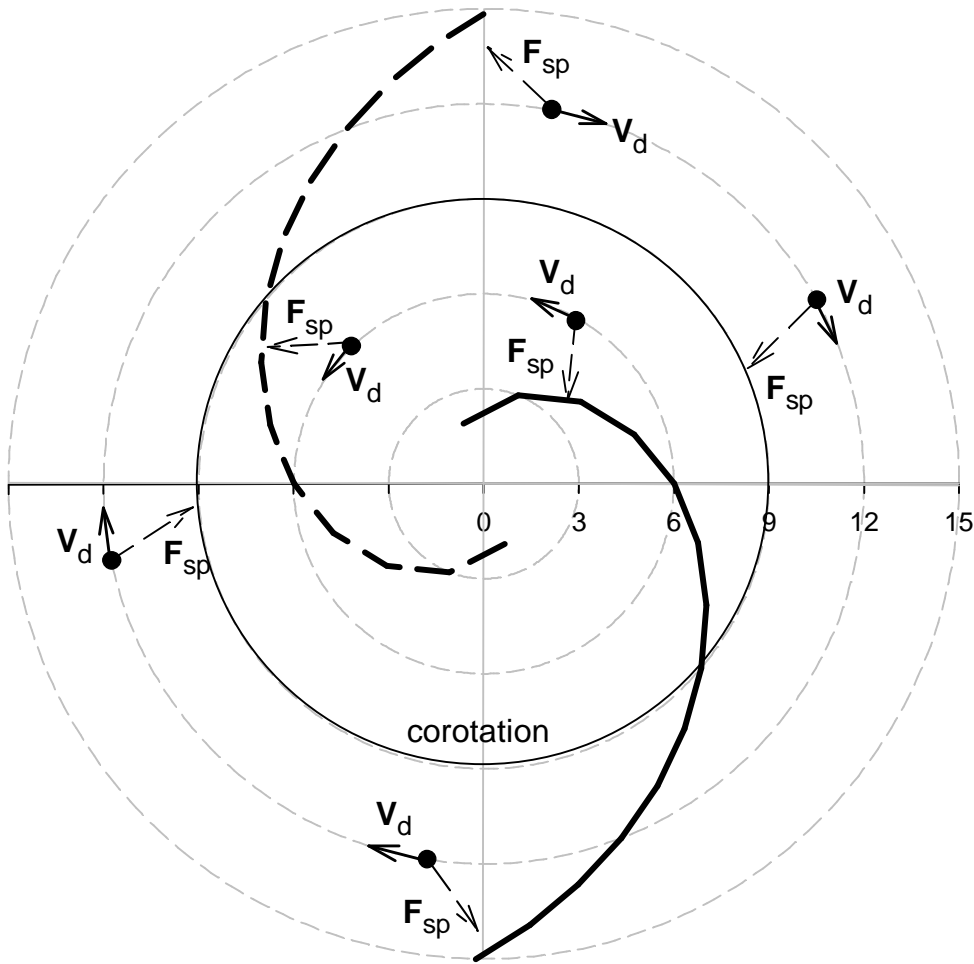}}
      \caption{Schematic description of gravity forces (shown by the dashed
      arrows) acting on test dust particles
      from the stellar spiral arms (plotted by thick solid and dashed lines).
      Velocities of dust particles in the
      frame of reference of stellar spiral arms are indicated
      by solid arrows. Two test particles inside corotation and four
      test particles outside corotation are shown. Arrows are
      not to scale.}
         \label{fig6}
\end{figure}

The phase ($r_{\rm d},\theta_{\rm d}$) trajectories of three representative dust particles
{\it in the frame of reference of stellar spiral arms}
are shown in Fig.~\ref{fig5}. The stellar spiral arms in Fig.~\ref{fig5}
are always located along azimuthal angles $\theta_{\rm sp}=0$ and $\pi$.
The dashed line shows the trajectory of a dust particle placed initially at
$r_{\rm d}$=5.0~kpc in the arm ($\theta_{\rm d}=0$; hereafter, particle~1),
whereas the solid line gives the trajectory of a dust
particle placed initially at $r_{\rm d}$=7.0~kpc in the interarm region
($\theta_{\rm d}=\pi/4$; hereafter, particle~2).
Dust particles similar to particle~1 may be injected by supernova
explosions which tend to occur near spiral arms.
Particle~2 may be created by red giant stars.

It is apparent that the trajectories of particles~1 and 2
are quite different. Particle~1 moves on an elliptical orbit in one direction
(counterclockwise),
remaining always inside the corotation circle ($r_{\rm cr}\approx 8.0$~kpc).
The trajectory of particle~1 is fairly stable.
It comes closer to the origin when traveling between the arms
($\theta_{\rm d} \approx \pi/2$) and it passes through the spiral arms
4~times during 0.65~Gyr. Since dust particles are thought to be destroyed
by supernova explosions, such a frequent encounter with the spiral arms
implies that particle~1 may not live longer than 0.5~Gyr (although
it depends on the covering factor of sufficiently strong shock waves,
$v_{\rm SN} \ga 100$~km~s$^{-1}$, generated by supernova explosions (SNe) and propagating
perpendicular to the spiral arms).
 Particle~2 has a stable, kidney-shaped trajectory, which shows an impressive
 radial migration of $\sim 4$~kpc, as opposed to only 1~kpc for particle~1.
Interestingly, particle~2 is caught between the spiral arms,
periodically passing through the corotation circle from the inner part of the
disk to the outer part and vice versa. Since particle~2 never encounters the spiral arms
during 2~Gyr, it may live significantly longer than particle~1.
Dust particles that are born at $\theta_{\rm d}\approx \pi/2$ (i.e. exactly
between the spiral arms) but are sufficiently away from the galactic center,
move on much more compact orbits as shown by the dotted line in Fig.~\ref{fig5}.

To summarize, we find that all particles born at $r<6$~kpc,
i.e. $\ga 2.0$~kpc away from the position of corotation radius $r_{\rm cr}\approx 8.0$~kpc,
always stay inside corotation. This is irrespective of their azimuthal
angles of birth.
Trajectories of those
dust particles resemble that of particle~1 shown in Fig.~\ref{fig5} by
the dashed line.
Such particles often encounter spiral arms and thus may be destroyed in one rotation
period by shocks from SNe.
The fraction of surviving dust grains
is determined by the probability for them to meet  a sufficiently
strong ($v_s\ga 100$ km s$^{-1}$) destroying shock wave inside the arm.
On the other hand, particles injected near corotation at $6~{\rm kpc}\la
r<8.0$~kpc
(and sufficiently away from spiral arms)
spend most of their lifetime in the interarm region. Therefore, most of these
dust particles can survive over the entire migration
period in the outer galactic regions.

\begin{figure*}
 \centering
  \includegraphics[width=14 cm]{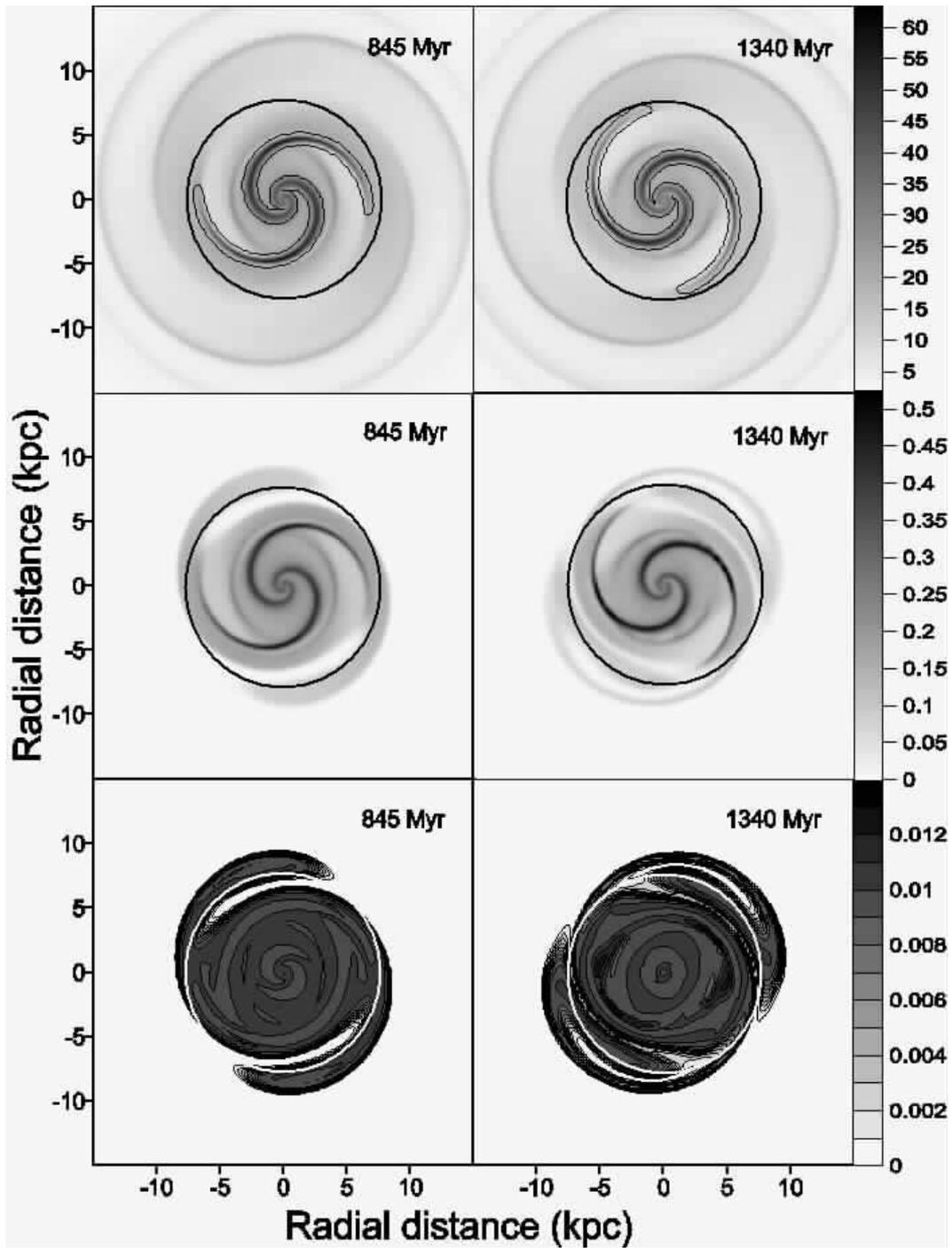}
      \caption{Gray-scale images of the gas distribution (the upper panels),
      dust distribution (the middle panels) and the dust-to-gas ratio (lower
      panels). The elapsed time is indicated in each panel.
      The position of corotation of the gas disk is outlined by
      the circle. The scale bars in the upper and middle panels give the column density in
      $M_\odot$~pc$^{-2}$}
         \label{fig7}
\end{figure*}

In the absence of spiral arms, dust particles move on circular orbits
determined by the axisymmetric global gravitational potential of a galaxy
and the drag force from the gas disk.
When the non-axisymmetric spiral gravitational potential is introduced,
dust particles move on various non-circular orbits as shown in Fig.~\ref{fig5}.
In order to understand the origin of such non-circular orbits, we consider
the influence of a non-axisymmetric spiral gravity force
on the motion of dust particles.
The gravity force of spiral arms acting on moving particles
is schematically plotted in Fig.~\ref{fig6} by the dashed arrows.
Six particles are presented; two particles inside the
corotation circle and four particles outside the corotation circle.
The direction of motion of dust particles {\it in the frame of reference
of the spiral pattern} is shown by the solid arrows.
Two spiral arms are schematically shown by the solid and dashed
lines. A moving dust particle may see the nearest spiral arm
as either convex or concave.
If the dust particle is closer to the concave spiral,
the latter acts to increase (on average) the radial distance of the particle
due to its gravitational drag.
On the contrary, if the dust particle is closer to the convex spiral, the latter
tends to decrease the radial distance of the particle.
If we now recall that the trailing spiral arms in Fig.~\ref{fig5} are located
along the azimuthal angles 0 and $\pi$ and rotate counterclockwise,
then the effect of gravitational drag from the spiral arms becomes
clearly evident in the trajectories of dust particles.
For instance, when dust particles are in the first quadrant, they are
attracted by the convex spiral located at $\theta_{\rm sp}=0$. As a consequence,
the radial position  of these particles decreases. When dust particles
enter the second quadrant, their motion is controlled by the concave spiral
located at $\theta_{\rm sp}=\pi$ and their radial position increases.
Therefore, the trajectories of dust grains may be
sensitive to the pitch angle of stellar spiral arms. The maximum radial
migrations of dust grains are expected for a pitch angle of $45^\circ$.

The action of the friction force is more difficult to assess.
The friction between
dust and gas may help reverse the direction of motion of a dust particle (with
respect to the spiral arms) when it passes through corotation.
Indeed, when the dust particle is inside the corotation circle, its velocity is
larger than that of spiral arms (the same is true for the gas in general).
After passing through corotation, the dust particle
enters the region which is characterized by a smaller velocity than that
of spiral arms. Therefore, we may expect that the friction between
gas and dust would act to equalize the velocities of dust and gas,
and thus decelerate the dust particles. This would eventually reverse
their direction of motion with respect to the spiral arms as seen
in Fig.~\ref{fig5} for particle~2.

\section{Response of dust to the stellar spiral density wave}
\label{dust}

In this section we study how dust responds to the imposed
spiral disturbance. The dominant sources of dust formation in disk galaxies
are stellar winds from late-type giant and supergiant stars, supernova explosions, and growth
of pre-existing grains in dense $n\ga 10^3$~cm$^{-3}$ molecular clouds
\cite{Dwek}. All of these sources are mainly located inside
the corotation circle. Indeed, the strongest gas response to the stellar
spiral density wave in Fig.~\ref{fig4} is obviously seen inside corotation.
We compute the critical density for star formation as defined by the Toomre criterion
\cite{Ken2}
\begin{equation}
\Sigma_{\rm crit}^{\rm \: Toomre} = {0.7\:c_{\rm s}\: \kappa \over \pi \:G},
\end{equation}
where $\kappa$ is the epicyclic frequency,
and the shear criterion \cite{Martin}
\begin{equation}
\Sigma_{\rm crit}^{\rm \: shear}={2.5 A_{\rm sh} \: v_{\rm s} \over \pi G},
\end{equation}
where $A_{\rm sh}=-0.5\:r\:{\rm d}\Omega /{\rm d} r$  is the local shear rate (the Oort constant).
The contour line in Fig.~\ref{fig4}
traces the region of supercritical density with $\Sigma_{\rm g}>\min(\Sigma_{\rm
crit}^{\rm \: Toomre}, \Sigma_{\rm crit}^{\rm \: shear})$ which is evidently
localized inside corotation.
Although the gas density outside corotation may occasionally become supercritical
(not shown in Fig.~\ref{fig4}) especially along the outer diffuse gas spirals,
a lower gas density and higher temperature as compared to the inner region
can work against star formation
in that region. We thus conclude that most of star formation in our model galaxy
is localized within the corotation circle.
The radial scale length of our
old stellar disk is $r_{\rm s}=3$~kpc which also implies that most of low to
intermediate mass stars that are capable of producing dust are confined
within the corotation circle.

We instantaneously inject $3.4\times 10^7~M_\odot$ of dust mass inside the corotation circle
($r<7.8$~kpc) and then determine the fraction of dust that can move outside corotation.
In the present simulations, the dust particles are injected at $t=0$~Myr
when a spiral structure in the gas disk has not yet developed. We also tried
to introduce dust at t=800~Myr (when the gas spirals are fully developed)
and confirmed that the time of dust injection has little influence on dust
dynamics.
The spatial distribution of injected dust is such that the dust-to-gas ratio
$\gamma=0.01$ inside corotation, and $\gamma=0$ outside corotation. The velocity of injected dust is equal to
that of the gas.
A conservative estimate of dust-to-gas mass ratio in the solar neighborhood
gives an average value of $\gamma=0.006$ ranging from $0.002$ up to $0.04$ in H$_2$
regions \cite{Spitzer}.

\begin{figure}
  \resizebox{\hsize}{!}{\includegraphics{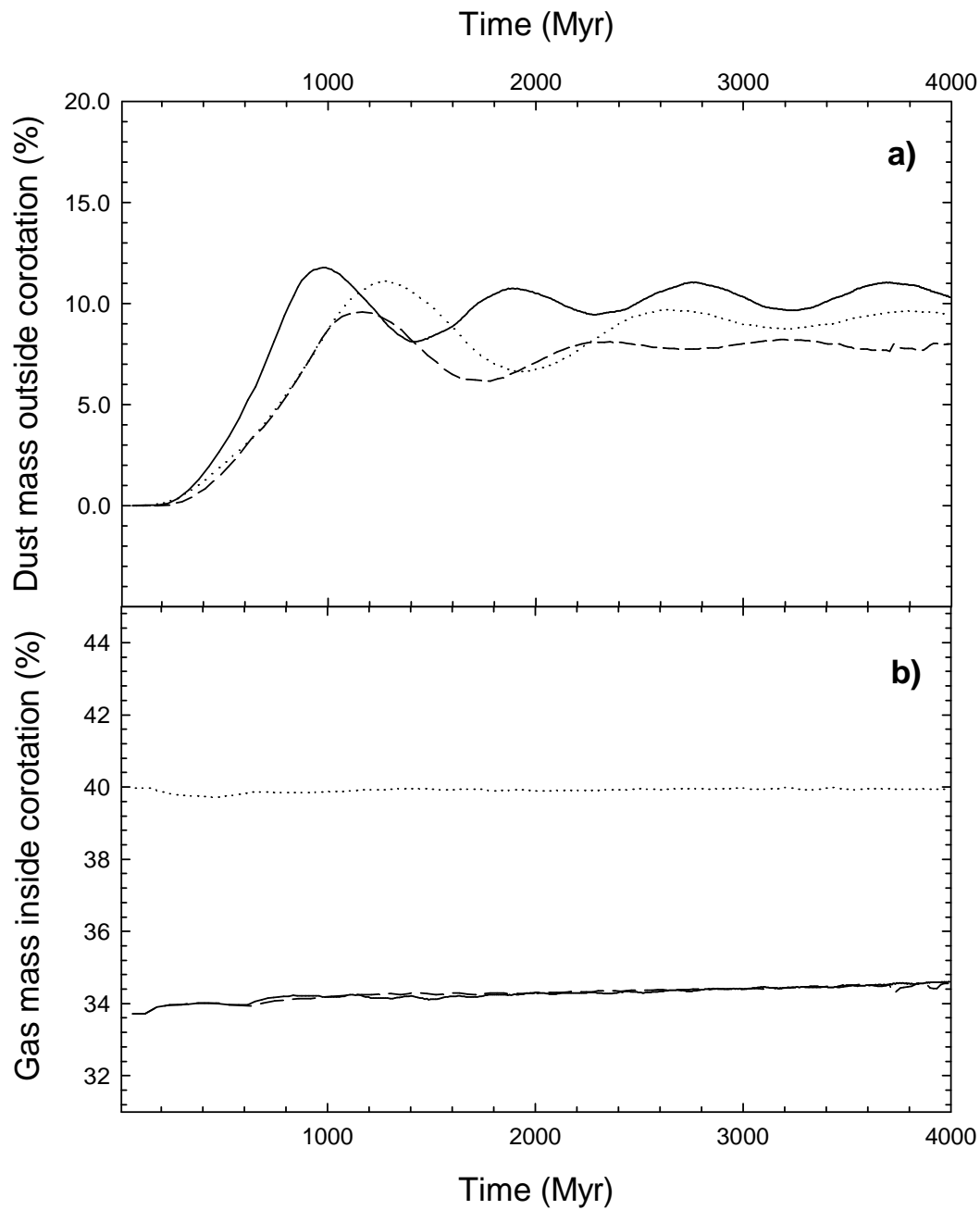}}
      \caption{ {\bf a)} Dust mass ouside corotation as a percentage of the total dust
      mass. The solid and dashed lines correspond to the models with pitch
      angles of stellar spiral density wave $i=25^\circ$ and $i=15^\circ$, respectively.
      The dotted line gives the dust mass outside corotation for a
      lower rotation frequency of stellar spirals $\Omega_{\rm sp}=21$~km~s$^{-1}$~kpc$^{-1}$.
      {\bf b)} Gas mass inside corotation as a percentage of the initial
      total gas mass. The solid, dashed, and dotted lines correspond to
      the same models as in the upper panel. The difference in the values
      of the gas fraction inside corotation for $\Omega_{\rm sp}=21$~km~s$^{-1}$~kpc$^{-1}$
      and $\Omega_{\rm sp}=24$~km~s$^{-1}$~kpc$^{-1}$ is due to the different
      positions of the corotation radius in the two cases.}
         \label{fig8}
\end{figure}

The upper and middle panels in Fig.~\ref{fig7} show two snapshots of the
evolution of the gas and dust surface densities, respectively,
whereas the lower panels give the dust-to-gas mass ratio.
The numbers in each panel indicate the elapsed time since the beginning of the
simulation. The circle indicates the position of corotation of the gas
disk. A strong spiral response to the stellar density wave develops
in the dust disk after approximately 3 revolution periods of our model galaxy.
The dust density closely follows that of gas in the inner regions of our model galaxy.
Small radial and azimuthal variations of the dust-to-gas ratio around the
initial value 0.01 are indicative of a dominant role of
frictional force in the dynamics of dust in the inner 5~kpc.
This is in agreement with the results of
Sect.~\ref{test} where we have found that dust particles born in the inner $5$~kpc
move on fairly stable elliptical orbits around the galactic center.
At larger radii $6~{\rm kpc}\la r<8.0$~kpc, however, significant variations of $\gamma$
become evident. There are regions near corotation that are completely devoid of
dust, even though the initial dust-to-gas ratio is 0.01.
A noticeable portion of dust is observed outside corotation. This dust tends to form into
diffuse outer spiral arms, which appear to be a smooth extension of the inner dust spiral
arms  and can be traced up to $r=10$~kpc.
The dust-to-gas ratio in the
outer dust arms is not dissimilar to that of the inner $r<5$~kpc region.
We thus conclude that the perturbing force of stellar spirals starts
to play an important role in the dust dynamics near corotation resulting
in an outward radial transport of dust.
If we now identify the regions of supercritical gas density
by contour lines in the upper panels of Fig.~\ref{fig7}, than it becomes
evident that the dust component can be found as far as 4-5~kpc from the sites of
active star formation.

The solid line in Fig.~\ref{fig8}a gives the dust mass outside corotation
(as a percentage of the total dust mass) versus time elapsed
since the injection of dust. Approximately $10\%$ of the initial
injected dust mass is transported outside corotation during 1~Gyr.
This implies that the dust transport rate is $\dot{M}=3.8~M_\odot$~Myr$^{-1}$.
At later times, the amount of transported dust varies
around $\approx 10\%$ of the total dust mass. These variations are mainly
due to dust particles born inside the corotation circle in the
interarm region, where they are trapped into periodic orbits as shown by
solid and dotted lines in Fig.\ref{fig5}.
We conclude that only a (small) fraction of dust
particles (approximately 10\% by mass) injected instantaneously inside
the corotation circle can move to the outer regions, while the rest is confined
in the inner galaxy. As a consequence, the radial transport of
an instantaneously injected portion of dust terminates after
$\approx 1$~Gyr.

The spiral stellar density wave considered so far has a pitch angle
of $i=25^\circ$. This is typical for Sc galaxies
\cite{Ken1}. The efficiency of outward radial transport of dust
may depend on the openness of stellar spiral arms. To test this conjecture,
we explored a series of models with  different pitch angles.
The dashed
line in Fig.~\ref{fig8} shows the the dust mass outside corotation as a
percentage of the total dust mass for the stellar spiral with a pitch angle $i=15^\circ$.
Roughly $8\%$
of the injected dust mass is pushed outside corotation, as compared to $\approx
10\%$
for $i=25^\circ$. This indicates that the efficiency of outward radial
transport of dust drops by roughly $2\%$.
We expect that the radial transport of dust by spiral
stellar density waves may becomes inefficient
for very tightly wound spirals with a pitch angle $i\la5^\circ$. Our simulations have also shown
that the efficiency of outward transport of dust is  weakly sensitive
to the angular velocity $\Omega_{\rm sp}$ of stellar spirals. For instance,
a stellar spiral with $\Omega_{\rm sp}=21$~km~s$^{-1}$~kpc$^{-1}$ can
transport roughly 1\% less dust (see the dotted line in Fig.~\ref{fig8})
than a spiral with $\Omega_{\rm sp}=24$~km~s$^{-1}$~kpc$^{-1}$.
We note that the corotation radius for $\Omega_{\rm sp}=21$~km~s$^{-1}$~kpc$^{-1}$
is at 8.8~kpc.

The solid line in Fig.~\ref{fig8}b shows the gas mass
inside corotation as a percentage of the total initial gas mass.\footnote{Due
to the imposed outflow outer boundary condition, we cannot reliably calculate the
gas mass outside corotation because its value can be affected
by the gas that leaves the computational area. The dust, however, never
reaches the outer boundary.} It is evident that the gas (contrary to the dust)
does not exhibit a noticeable radial migration. A very weak inward radial transport ($\sim1\%$
by mass) is seen for $\Omega_{\rm sp}=24$~km~s$^{-1}$~kpc$^{-1}$, and it
becomes negligible for  $\Omega_{\rm sp}=21$~km~s$^{-1}$~kpc$^{-1}$.
We attribute this difference between the gas and dust radial transport rates
to the restoring force of the pressure gradients
which are always present in the warm ($T_{\rm g}\sim 10^4$~K) gas.
The warm gas resists radial redistribution by creating
additional pressure gradients which act to restore the initial centrifugally
balanced configuration.
The dust, however, is cold and it can easily flow in the radial direction provided
that the coupling  between dust and gas is not very strong.

\begin{figure}
  \resizebox{\hsize}{!}{\includegraphics{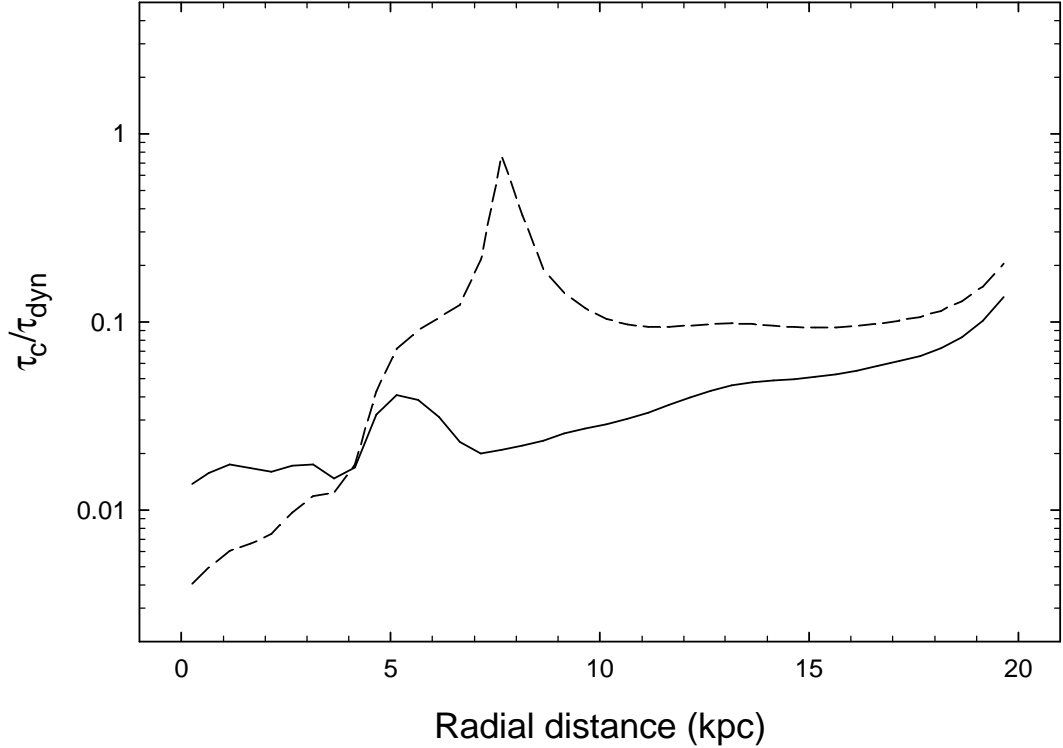}}
      \caption{The ratio of the frictional timescale $\tau_{\rm c}$ to
      the dynamical timescale $\tau_{\rm dyn}$ as a function of radius
      at $t=1340$~Myr.  The solid and dashed lines show  $\tau_{\rm c}/\tau_{\rm
      dyn}$ for the dynamical
      timescale defined as $\tau_{\rm dyn}=\Omega_{\rm d}^{-1}$ and
      $\tau_{\rm dyn}=|\Omega_{\rm d}^{-1} - \Omega_{\rm sp}^{-1}|$, respectively.
      The detailed explanation is given in the text.}
         \label{fig9}
\end{figure}

To determine the main cause of the radial transport of dust, we compute the
quantity  $\tau_{\rm c}/\tau_{\rm dyn}$  (averaged over 0.5~kpc radial annuli).
This quantity as a function of radius is shown by the solid line in Fig.~\ref{fig9} 
at the time when the gas spiral is fully developed.
The collisional timescale $\tau_{\rm c}$ of energy and momentum exchange
between the dust and gas particles is defined in Eq.~(\ref{thick}).
The local dynamical timescale is defined as $\tau_{\rm dyn}=\Omega_{\rm
d}^{-1}=r/v_{\rm d,\phi}$, where $v_{\rm d,\phi}$ is the azimuthal component
of dust velocity $\bl v_{\rm d}$.
If $\tau_{\rm c} \ll \tau_{\rm dyn}$ the dust is expected to be strongly
frictionally coupled to the gas and to essentially track the gas motion.
Figure~\ref{fig9} shows that $\tau_{\rm c}$ is a factor of $0.01-0.1$
smaller than $\tau_{\rm dyn}$, which implies a strong frictional coupling
betwen dust and gas.
However, judging from the bottom panels of Fig.~\ref{fig7}, it would appear that
the dust does not follow exactly the gas motion in the region near
corotation. The reason is that the dust particles have natural resonant
frequencies near corotation (remember that corotation determines the position
of resonance in gas/dust orbits).
If the gravitational field generated by spiral structure perturbs a dust
particle's orbit at or near its resonant frequency, then the responce of
the orbit will be large, even when the perturbing field is weak and/or
the frictional coupling is strong. To investigate this effect, we re-define
the local dynamical time as $\tau_{\rm dyn}=|\Omega_{\rm d}^{-1}-\Omega_{\rm
sp}^{-1}|$ and plot the corresponding quantity $\tau_{\rm c}/\tau_{\rm dyn}$ in Fig.~\ref{fig9} by
the dashed line. It is now evident that the frictional coupling between dust
and gas is much weaker near corotation. This is consistent with
the trajectories of dust particles found in Sect.~\ref{test} -- only the
dust particles that are born near corotation show substantial radial migration.

Another source of uncertainty in our simulations is the friction
coefficient $A$,  which we have adopted in the form of Eq.~(\ref{thick})
to simplify the computations. As is pointed out in Sect.~\ref{basic},
the total stellar+gas surface density $\Sigma_{\rm s}+\Sigma_{\rm g}$ (instead of simply
gas surface density, $\Sigma_{\rm g}$) should appear in the definition of the gas
vertical scale height
\begin{equation}
z_0={c_{\rm s}^2 \over \pi G (\Sigma_{\rm s}+\Sigma_{\rm g})},
\label{height}
\end{equation}
which may considerably increase the value of $A$. 
The uncertainity in the value of $\Sigma_{\rm s}$ (we do not explicitly
follow the evolution of the stellar disk) works against this re-definition
of $z_0$ in our simulations. 
In order to estimate the effect that the increased surface density of gravitating
matter in Eq.~(\ref{height}) may have on the radial transport of dust particles,
we notice that in the region of interest near corotation the ratio of stellar-to-gas
surface densities is usually in the $1.5-15$ range.
Therefore we performed a test run in which $A$ was defined by Eq.~(\ref{thick}),
but was later multiplied by a factor of 10.
The efficiency of outward radial transport of dust increased by less than $1\%$.
In another test, $A$ is decreased by a factor
of 10, and the efficiency of outward radial transport of dust is dropped by approximately
$1\%$. We conclude that an order of magnitude variation in $A$ has a
minor effect on the efficiency of radial transport of dust.
This reinforces a previous conclusion: the resonant region
near corotation, where $\Omega_{\rm d}\approx\Omega_{\rm sp}$, and the collisional drag 
force is dynamically less important than the spiral gravitational field, determines the radial
migration of dust.

\section{Dust destruction}
In this section we study the effect of dust destruction by SNe on the efficiency
of radial transport of dust in spiral galaxies.
The dust grain destruction by SNe is the most important
mechanism for cycling the dust back to the gas phase. The dust destruction
rate can be determined as \cite{Dwek}
\begin{equation}
D_{\rm d}(r,\phi,t)=m_{\rm dest}(r,\phi,t) \: \nu(r,\phi,t).
\label{destr}
\end{equation}
Here, $\nu(r,\phi,t)$ is the supernova rate per unit area and $m_{\rm dest}(r,\phi,t)$
is the mass of dust that is destroyed by a single SN expanding at location
$(r, \phi)$ and time $t$. Only SN remnants with velocities exceeding
100~km~s$^{-1}$ can efficiently destroy dust grains. Therefore, the value
of $m_{\rm dest}(r,\phi,t)$ is usually estimated from the relation $m_{\rm
dest}/M_{\rm d}= \epsilon M_{\rm g}^{100}/M_{\rm g}$, where $M_{\rm g}^{100}=3.8\times
10^3~M_\odot$ \cite{LF}
is the mass of a remnant accelerated to velocities $>100$~km~s$^{-1}$ by the blast.
This relation simply states that the fraction of destroyed
dust (by mass) within a computational cell is directly proportional to the
fraction of accelerated remnant mass. The
efficiency of dust destruction is $\epsilon \sim 0.1$  \cite{McKee}.

The SN rate per unit area  is defined as
\begin{equation}
\nu(r,\phi,t)={\int \limits_{9M_\odot}^{m_{\rm up}} \:  \Sigma_{\rm SFR}(r,\phi,t-\tau(m)) \: m^{-\alpha}
\: dm \over  \int \limits_{m_{\rm low}}^{m_{\rm up}} m^{1-\alpha}
dm },
\label{dustsource}
\end{equation}
where $\Sigma_{\rm SFR}$ is the star formation rate per unit area, $\alpha$ is the
slope of initial mass function (IMF), $\tau(m)$ is the lifetime of a star of
mass $m$, $m_{\rm low}$ and $m_{\rm up}$
are the lower/upper cutoff masses, respectively.
In practice, we approximate the integral in Eq.~(\ref{dustsource}) by the
sum with a mass discritization $dm=1~M_\odot$.

Observations of both normal and starburst disk galaxies
suggest that on the scales of a few kiloparsecs star formation may be
represented by a Schmidt law  \cite{Ken3}
\begin{equation}
\Sigma_{\rm SFR}\: \left({M_\odot \over {\rm yr}~{\rm kpc}^{2}}\right)=2.5\times 10^{-4}~\Sigma_{\rm g}^{\;1.5}
\:\left({M_\odot\over{\rm pc}^2}\right).
\label{Schmidt}
\end{equation}
We modify the star formation law (\ref{Schmidt}) by assuming that star formation
is suppressed when $\Sigma_{\rm g} < min(\Sigma_{\rm crit}^{\rm \: Toomre},
\Sigma_{\rm crit}^{\rm \; shear})$.

Often, the instantaneous recycling approximation is used to solve  Eq.~(\ref{destr}),
which neglects the time delay that enters Eq.~(\ref{destr}) through the SN
rate $\nu$.
However, we follow a more accurate approach and form
stellar clusters  each Myr according to the star formation law (\ref{Schmidt}).
Stellar clusters are assigned positions and velocities
drawn from the parent gas and are evolved as collisionless particles
in the combined gravitational potential of the halo and stellar
disk. The local self-gravity among stellar clusters is neglected.
Since each stellar cluster carries information on the star formation rate at the time of its
birth, the supernova rate $\nu(r,\phi,t)$ and the dust destruction rate
$D_{\rm d}(r,\phi,t)$ in each computational cell $(r,\phi)$
at a time $t$ can be easily obtained. This approach was successfully applied
by Vorobyov (\citeyear{Vor}) to model the $H\alpha$ luminosity in ring galaxies.
We use the Salpeter IMF with $\alpha=2.35$ and lower/upper cutoff masses $m_{\rm low}=0.5~M_\odot$
and $m_{\rm up}=40~M_\odot$, respectively.

We modify the hydrodynamic equations (\ref{cont})-(\ref{second}) of our gas+dust system
to take into account star formation and dust destruction.
The continuity equations for the gas and dust become
\begin{equation}
{\partial \Sigma_{\rm g} \over \partial t}+ {\bl \nabla} \cdot (\Sigma_{\rm
g}\, {\bl v}_{\rm g})=S_{\rm g}-D_{\rm g},
\label{contgas}
\end{equation}
\begin{equation}
{\partial \Sigma_{\rm d} \over \partial t}+ {\bl \nabla} \cdot (\Sigma_{\rm
d}\, {\bl v}_{\rm d})=-D_{\rm d}.
\label{contdust}
\end{equation}
The resulting momentum equations are
\begin{eqnarray}
{\partial \over \partial t} \, \Sigma_{\rm g} {\bl v}_{\rm g}  +({\bl v}_{\rm g} \cdot {\bl
\nabla}) \Sigma_{\rm g} {\bl v}_{\rm g} &=& -\Sigma_{\rm g}{\bl \nabla} \Phi_{\rm s1,s2,h} - {\bl \nabla} P_{\rm
g}  +\Sigma_{\rm d} {\bl f} + \nonumber \\
&&{\bl v}_{\rm g} (S_{\rm g}- D_{\rm g}),
\label{motion1} \\
{\partial \over \partial t} \, \Sigma_{\rm d} {\bl v}_{\rm d}  + ({\bl v}_{\rm d} \cdot {\bl
\nabla}) \Sigma_{\rm d} {\bl v}_{\rm d} &=& - \Sigma_{\rm d} {\bl \nabla} \Phi_{\rm s1,s2,h}
-\Sigma_{\rm d} {\bl f} - {\bl v}_{\rm d} D_{\rm d}.
\label{motion2}
\end{eqnarray}
We have rewritten equations of motion in the form of
momentum equations for $\Sigma \,\bl v$ which are appropriate for the system with sources
and sinks.
Star formation depletes the gas reservoir of our model galaxy and is taken into
account by the sink term $D_{\rm g}=\Sigma_{\rm SFR}$ in Eqs.~(\ref{contgas})
and (\ref{motion1}).
The gas is returned to the system with supernova explosions and quiet mass
loss of intermediate  and low mass stars.
Since we do not use a multiphase description of the interstellar medium,
the ejected hot gas is directly transformed into a warm ($T\sim 10^4$~K) phase.
The rate of gas ejection per unit area is determined by
\begin{equation}
S_{\rm g}={\int \limits_{m_{\rm low}}^{m_{\rm up}} g(m) \:  \Sigma_{\rm SFR}(r,\phi,t-\tau(m)) \: m^{-\alpha}
\: dm \over  \int \limits_{m_{\rm low}}^{m_{\rm up}} m^{1-\alpha}
dm },
\label{gassource}
\end{equation}
where $g(m)$ denotes the gas mass ejected by a star of mass $m$ \cite{KA}.
We found it computationally prohibitive to compute accurately (i.e. by taking
into account the time delay $\tau(m)$ in Eq.~[\ref{gassource}]) the gas ejection
rate by stellar clusters on a time scale of interest (a few Gyr).
Therefore, for the stellar clusters older than $100$~Myr, the instantaneous
recycling approximation is assumed by neglecting the time delay in Eq.~(\ref{gassource}).
These stellar clusters instantaneously release all gas that can be produced
by the $0.5 - 5~M_\odot$ stars.
Further dynamical evolution of such clusters is not computed.

\begin{figure}
  \resizebox{\hsize}{!}{\includegraphics{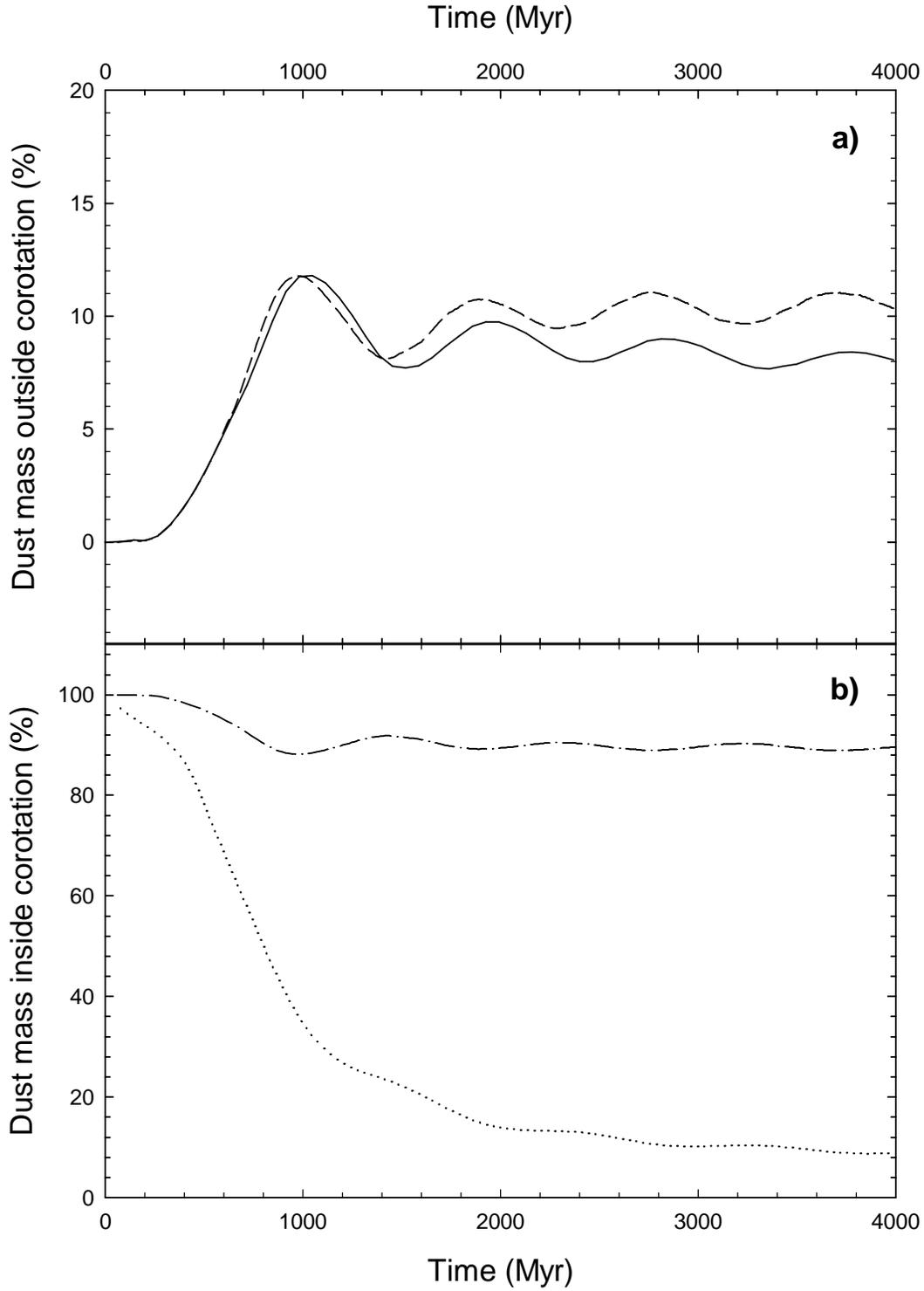}}
      \caption{{\bf a)} Dust mass outside and {\bf b)} inside corotation as a percentage of
      the total initial dust mass.
      The solid and dashed lines show the relative dust mass outside corotation in the models with and without
      dust destruction, respectively. The dotted and dotted-dashed lines
      give the relative dust mass inside corotation in the models with
      and without dust destruction, respectively.}
         \label{fig10}
\end{figure}

We study the effect of dust destruction on the radial transport of dust
using a prototype model in which the spiral density wave has a pitch angle $i=25^\circ$
and angular velocity $\Omega=24$~km~s$^{-1}$~kpc$^{-1}$.
This model was studied in detail in Sect.~\ref{dust} where we neglected the
effect of dust destruction. It was shown there that about $10\%$ of dust
by mass injected instantaneously inside the corotation circle can be
transported outside this region by
the gravitational drag associated with spiral density waves.
We reproduce this result in Fig.~\ref{fig10}a by the dashed line, and
the solid line gives the dust mass outside corotation as a percentage of the initial
total dust mass when dust destruction is taken into account.
As is seen, supernova explosions have a minor effect on the
radial transport of dust, decreasing the relative amount of dust outside
corotation by only $2\%$ when compared to the model without dust destruction.
It is worth noting that inside corotation the destruction of dust by supernova
explosions is efficient. Indeed, the dashed-dotted and dotted lines in Fig.~\ref{fig10}b
show the dust mass inside corotation as a percentage of the initial total
dust mass when dust destruction is taken and not taken into account, respectively.
Approximately $15\%$ of the instantaneously injected dust mass is left
inside corotation after 2~Gyr. Given that another $10\%$ has been
transported outside corotation, SNe have destroyed about $75\%$ of the
injected dust mass during 2~Gyr. In Sect.~\ref{dust} we argued that
the ultimate fate of dust grains would depend on their place of birth.
The simulations discussed in this section reinforce our earlier conclusion.
In our model dust grains are formed inside the corotation circle at $r_{\rm
cr}\approx 8$~kpc.
The dust grains that are born
in the inner 6.0~kpc circle (by implication they constitute the majority
of the total dust mass) move on elliptic orbits periodically crossing
the spiral arms and will be inevitably destroyed by supernova explosions
after several revolution periods.
However, the dust grains that are born in the $6~{\rm kpc}\la r <8$~kpc annulus and
sufficiently away from spiral arms (by implication they constitute
the minority of the total dust mass) manage to survive the destructive effect
of supernova explosions during many revolution periods  of our model galaxy.
These dust grains  may move on orbits similar to those shown in Fig.~\ref{fig5}
by the solid and dotted lines, circulating between the spiral arms (but never crossing
them) and periodically traveling in and out of the corotation circle.
We conclude that a (small) portion of
dust grains may survive the destructive
effect of SNe during many ($>10$) revolution periods of a spiral galaxy.

\section{Summary}

In this paper we study the outward radial transport of interstellar dust grains
which is connected with the perturbation of circular motions caused by spiral
stellar density waves in disk galaxies. We consider a model spiral galaxy
in which star formation (and, by implication, dust formation) is localized
within the corotation circle.
We find that the combined action
of  the gravity force of spiral stellar density waves and the drag force between the
gas and dust components imparts to dust particles a radial velocity component,
and thus can evacuate dust outward.
We conclude that:

1. Dust grains that are formed inside corotation can be transported to
a distance that exceeds the corotation radius by roughly $25\%$.
In particular, if corotation is located at $r_{\rm cr}=8.0$~kpc, dust grains formed
inside the $6~{\rm kpc}\la r<8.0$~kpc annulus can travel to distances as large as 10~kpc in roughly 1.0~Gyr.
This is approximately an order of magnitude faster that can be provided by the interstellar turbulence.

2. A fraction of dust grains can be trapped into the interarm region,
and thus can escape a hostile environment from supernova explosions acting
mostly in the spiral arms. These grains form diffuse spiral arms which
extend 4-5~kpc from the sites of active star formation. Dust particles that cross
the spiral arms during their drift outward can survive if they do not
meet strong shocks from SNe while passing through the arm.

3. In the model that considers instantaneous dust injection, about $10\%$ of dust by mass
is evacuated outside the corotation circle in roughly 1.0~Gyr. One can
assume that continuous injection will result in the same percentage of
evacuated dust. The efficiency of outward radial dust transport is weakly
sensitive to the rotation speed of stellar spirals, but it may be considerably reduced
for very tightly wound spirals with a pitch angle $i\la5^\circ$.
Dust destruction by supernova explosions has only a minor effect on the
efficiency of outward radial transport of dust by spiral density waves,
which
decreases the relative amount of transported dust mass by $\sim 2\%$.
A (small) portion of dust grains may survive the destructive
effect of SNe during many ($>10$) revolution periods of a spiral galaxy.

4. Our modeling shows that the dust-to-gas ratio in the region near corotation
($\pm 2$~kpc) can vary substantially over the scales
comparable to the thickness of spiral arms, $\sim 1$ kpc. Possible
observations of such variations would be suggestive of the proposed mechanism
of radial dust transport.

In our model we neglected the fact that charged dust grains are strongly
coupled to the magnetic field -- the gyration frequency for a typical grain
with radius $a\sim 0.1\mu$m and grain charge $Z\sim 100$ is of $\omega_B\sim
10^{-10}$ Hz. We expect that in more realistic models, where
the magnetic field is permanently regenerated by a turbulent dynamo, dust particles
will be involved in both the turbulent diffusive motions and regular
flow, and the overall picture will remain qualitatively similar to that
described above. We leave this issue for further study.

We should mention that the interaction of gas and solid particles
through the drag force and the associated radial flow
of solid particles have been already extensively studied in
the context of migration of planetisimals in protoplanetary
disks.  For example, Weidenschilling (\citeyear{WS}) has noticed that in
protoplanetary disks the planetesimals should migrate in the direction of increasing
pressure (i.e. inward, for an axisymmetric disk with a declining radial density
profile). This is due to the fact that the gas in protoplanetary disks
generally
orbits the central protostar at the sub-Keplerian velocity due to a substantial contribution
of pressure gradients in the gas rotation curve. The cold planetesimals (which by
implication have no pressure)  are then
forced to flow inward if the frictional orbital coupling between the gas and planetesimals
is not too strong. This mechanism however may not work on galactic scales because
the contribution of the pressure gradients to the gas rotation velocity
may be negligible (see Sect.~\ref{equilib}). In fact, the pressure gradients may even
assist the gravity in some disk galaxies with a ring-like radial
distribution of gas. Another interesting aspect of gas-planetesimal interaction
in self-gravitating, protoplanetary disks was investigated by Rice et al.
(\citeyear{Rice}). They considered disks with a flocculent spiral structure
and demonstrated that the drag force rather than gravitational field of
spiral arms plays a dominant role in the dynamics of planetesimals of intermediate
size, forcing them to concentrate in the gas spiral arms.
The non-axisymmetric gravitational field of grand-design spirals
(considered in the present paper) is by implication much stronger than
that of flocculent spirals. As a consequence, the trajectories of dust
particles can be considerably influenced by the spiral gravitatioanl field
(Fig.~\ref{fig5}). We believe that the resonant action of spiral gravitational
field on dust particles (rather than the drag force between the gas and dust)
is the main driving force for the radial
outward transport of dust in grand-design spiral galaxies. This conjecture is justified by the fact that
the radial flow of dust takes place near the corotation resonance where 
a weaker frictional coupling  between the dust and gas occurs (see Fig.~\ref{fig9}).

\section*{Acknowledgements}
      We are grateful to Dr. Carol Jones for carefully reading the manuscript
      and correcting the English language usage. We are thankful to the
      anonymous referee for an insightful report that helped improve the
      final presentation.
       EIV gratefully acknowledges present support from a CITA National Fellowship.
      Part of this work (YS) was supported by
      \emph{Deut\-sche For\-schungs\-ge\-mein\-schaft, DFG\/}
      (project SFB TP B3). The simulations were partly performed on the
      Shared Hierarchical Academic Research Computing Network (SHARCNET) cluster.

\end{document}